\documentclass[preprint,prb,showpacs,citeautoscript]{revtex4}

\usepackage{graphicx}

\begin{document}

\title{Suppressed Magnetization in
La$_{0.7}$Ca$_{0.3}$MnO$_3$/YBa$_2$Cu$_3$O$_{7-\delta}$ 
Superlattices}

\author{A. Hoffmann}
\email{hoffmann@anl.gov}
\author{S. G. E. te Velthuis}
\affiliation{Materials Science Division, Argonne National Laboratory, 
Argonne, Illinois 60439} 
\author{Z. Sefrioui}
\author{J. Santamar{\'\i}a}
\affiliation{GFMC. Dpto. Fisica Aplicada III, Universidad Complutense
de Madrid, 28040 Madrid, Spain}
\author{M. R. Fitzsimmons}
\author{S. Park}
\affiliation{Los Alamos National Laboratory Los Alamos, New Mexico
87545}
\author{M. Varela} 
\affiliation{Condensed Matter Sciences Division, Oak Ridge National Laboratory,
Oak Ridge, Tennessee 37831-6030} 

\date{\today}

\begin{abstract}
  We studied the magnetic properties of 
La$_{0.7}$Ca$_{0.3}$MnO$_3$ / YBa$_2$Cu$_3$O$_{7-\delta}$ superlattices.  
Magnetometry showed that with increasing YBa$_2$Cu$_3$O$_{7-\delta}$ layer thickness 
the saturation magnetization per La$_{0.7}$Ca$_{0.3}$MnO$_3$ layer decreases.  From 
polarized neutron reflectometry we determined that this magnetization reduction is due to an 
inhomogenous magnetization depth profile arising from the suppression of magnetization 
near the La$_{0.7}$Ca$_{0.3}$MnO$_3$ / YBa$_2$Cu$_3$O$_{7-\delta}$ interface.  
Electron energy loss spectroscopy indicates an increased 3d band occupation of the Mn atoms 
in the La$_{0.7}$Ca$_{0.3}$MnO$_3$ layers at the interface.  Thus, the suppression of 
ferromagnetic order at the La$_{0.7}$Ca$_{0.3}$MnO$_3$ / 
YBa$_2$Cu$_3$O$_{7-\delta}$ interface is most likely due to charge transfer between the 
two materials.
\end{abstract}


\maketitle

  The interplay between ferromagnetism and superconductivity has been of longstanding 
research interest, since the competition between these generally mutually exclusive types of 
long-range order gives rise to a rich variety of phenomena\cite{SC/FM-Review}.  
Using ferromagnetic (F)/superconducting (S) layered heterostructures enables investigation 
of diverse effects, i.e., non-monotonic changes of the superconducting transition 
temperature\cite{Sam} $T_c$ and $\pi$-Junctions\cite{pi-junction} in S/F/S structures, 
and the dependence of $T_c$ on the relative magnetization direction in F/S/F 
structures\cite{SC-spin-valve}.  Layered heterostructures also offer opportunities to help 
resolve theoretical predictions with respect to F/S structures, such as triplet 
superconductivity\cite{triplet}.

  Most studies of F/S heterostructures involved transition metal systems.  But there is 
increasing interest in F/S structures in which combinations of complex materials based on 
perovskite oxides are used, since these materials on their own show unusual properties.  
Many different cuprate high-$T_c$ superconductors are characterized by a short 
superconducting coherence length and an anisotropic superconducting gap.  At the same time 
manganites are atypical ferromagnets in that they exhibit colossal magnetoresistance and are 
potentially half-metallic.  Since both classes of materials have very similar perovskite 
structures with comparable lattice constants in the basal plane, it is possible to combine them 
into structurally coherent superlattices with very sharp interfaces\cite{Pedro,Habermeier}.  
Previous experiments have already shown that cuprate/manganite based superlattices have 
distinctively different properties compared to their transition metal counterparts.  They 
exhibit unusually long ranging proximity effects\cite{long-proximity}, spin injection into 
the superconducting layer\cite{spin-diffusion}, and even giant 
magnetoresistance\cite{new-GMR}.

  Previous work mostly focused on the influence of proximity effects on the 
superconducting properties like the superconducting transition temperature or critical 
current density.  Here, we focus on the magnetic properties of the magnetic layers.  In this 
paper we show that in La$_{0.7}$Ca$_{0.3}$MnO$_3$ (LCMO) / 
YBa$_2$Cu$_3$O$_{7-\delta}$ (YBCO) superlattices the ferromagnetic ordering is 
suppressed at the interface between LCMO and YBCO.  This adds more complexity for 
explaining proximity effects in oxide based heterostructures.

  The LCMO/YBCO superlattices were grown by high pressure sputtering on (100) 
SrTiO$_3$ substrates.  For one series of superlattices the LCMO layer thickness was kept 
constant at 15 unit cells (corresponding to 60~\AA), while the YBCO layer thickness was 
varied between 1 and 12 unit cells.  X-ray diffraction indicated that the samples are epitaxial 
and x-ray reflectivity showed that the interfaces are well defined with roughnesses below one 
unit cell for the layers close to the substrate.  The roughness increased to about 35~\AA\ for 
the top-most layers.  Scanning tranmission electron microscopy also indicated a similar trend 
for the interface roughnesses.  Magnetic hystersis loops were measured with a Quantum 
Design superconducting quantum interference device system.  Polarized neutron 
reflectometry measurements were taken on $5 \times 10$~mm$^2$ samples using POSY1 at 
IPNS (Argonne) and ASTERIX at LANSCE (Los Alamos).  Electron microscopy 
observations and electron energy loss spectroscopy (EELS) measurements were obtained in a 
dedicated scanning transmission scanning microscope VG Microscopes HB501UX, operated 
at 100~kV and equipped with a Nion aberration corrector and an Enfina EEL spectrometer.  
Cross sectional specimens were prepared by conventional methods: grinding, dimpling, and 
Ar ion milling at 5~kV.  Final cleaning was performed at 0.5~kV.

\begin{figure}
\includegraphics[width=3.4in]{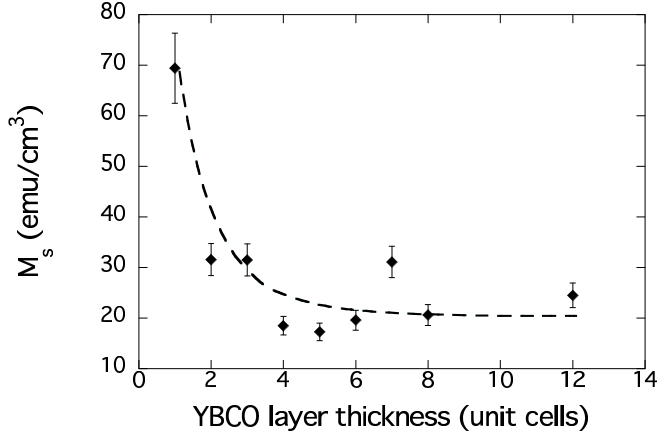}
\caption{Saturation magnetization measured at 100~K normalized to the LCMO volume 
for [LCMO 15 u.c./YBCO $n$ u.c.] superlattices with varying YBCO thickness.  The dashed 
line is a guide to the eye.}
\label{fig:magnetization}
\end{figure}

  The saturation magnetization attributed to the LCMO layers shows a drastic reduction with 
decreasing LCMO thickness (Fig.~\ref{fig:magnetization}).  The saturation magnetization 
is well below both the bulk value\cite{bulk_LCMO} of $M_s = 576$~emu/cm$^3$ and the 
values of $M_s \approx 400$~emu/cm$^3$ observed for single layer LCMO thin 
films\cite{LCMO_film} prepared under similar conditions as the superlattices discussed in 
this paper.  Previously, the saturation magnetization of YBCO/LCMO superlattices were 
observed to achieve values comparable to single layer LCMO films only when the LCMO 
layer thickness exceeded 100 unit cells\cite{proximity} (in our case the thickness is 15 unit 
cells).  This observation implies that the proximity of YBCO to LCMO affects the magnetic 
properties of LCMO for either intrinsic or extrinsic reasons.

  An important key to understanding the origin of the reduced magnetization in the LCMO 
can be found in the magnetization depth profile of each LCMO layer.  This information can 
be readily obtained from polarized neutron reflectometry\cite{neutron-review}.  Briefly, 
the technique involves reflection of a polarized neutron beam with wave vector 
$\mathbf{k_i}$ from the sample onto a polarization analyzer with wave vector 
$\mathbf{k_f}$.  Use of a polarized beam with polarization analysis permits determination 
of the spin-dependent neutron reflectivities.  The difference between the non-spin-flip 
reflectivities $R^{++}$ and $R^{--}$ (with the neutron spin parallel and antiparallel to the 
to the applied field, respectively) is determined by the component of the magnetization 
depth profile, which is parallel to the applied magnetic field and perpendicular to 
$\mathbf{q}=\mathbf{k_f}-\mathbf{k_i}$.  In our experiment the magnetic field is 
applied along the surface plane.  Since we performed our measurements in saturation (as 
determined by magnetometry), the perpendicular magnetization component is zero and we 
only present non-spin-flip reflectivities.  The momentum transfer ($q$-dependence) of the 
reflectivities is related to the Fourier components of the magnetization depth profile -- 
providing depth sensitivity.\cite{Mike-Chuck}  Using an iterative process\cite{Parratt}, 
the reflectivities calculated from model structures comprised of chemical and magnetical 
depth profiles, can be fitted to the observed reflectivities.

\begin{figure}
\includegraphics[width=3.4in]{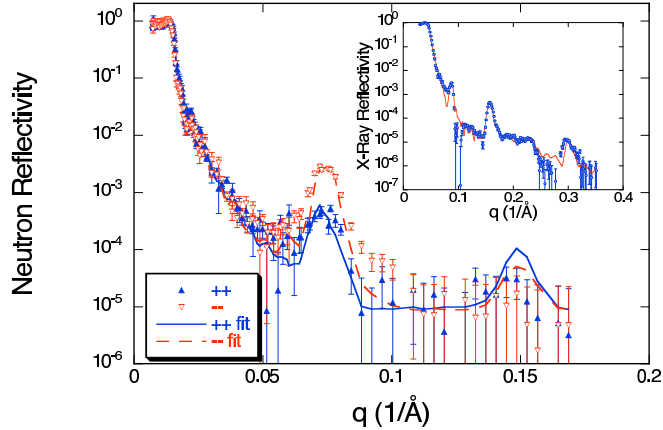}
\caption{(Color online) Polarized neutron reflectivity for the [LCMO 15 u.c/YBCO 3 u.c.] 
superlattice in saturation.  The inset shows X-ray reflectivity for the same sample.  The 
experimental data is presented by the symbols, while the lines are fits.  The fit for the 
neutron data is based on the model using an inhomogeneous magnetization as discussed in 
the text.}
\label{fig:PNR}
\end{figure}

  Figure~\ref{fig:PNR} shows polarized neutron reflectivities for a sample with 3 unit cell 
thick YBCO layers.  The measurements were taken with the sample at 120~K and an applied 
field of 5.4~kOe.  The first and second Bragg peaks due to the superlattice structure are 
clearly visible.  In order to analyze the neutron reflectivity data we first determined the 
chemical structure with X-ray reflectivity, which is shown with the corresponding fit in the 
inset of Fig.~\ref{fig:PNR}.  The structural parameters were then used for the subsequent 
fit of the neutron reflectivity data, where the only free parameters were the neutron 
scattering length densities and the magnetization depth profile.  

\begin{figure}
\includegraphics[width=3.4in]{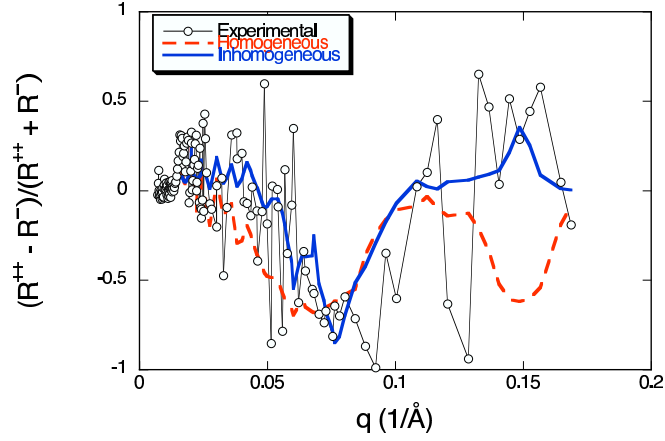}
\caption{(Color online) Spin asymmetry $(R^{++}-R^{--})/(R^{++}+R^{--})$ of the 
neutron reflectivity.  The line with symbols indicate the experimental data, while the dashed 
and solid line indicate the fitted spin asymmetry for the homogeneous and inhomogeneous 
models, respectively.}
\label{fig:Polarization}
\end{figure}

  For the fit of the polarized neutron reflectivity data we used two different models for the 
magnetization depth profile.  One where the magnetization is homogeneous throughout each 
LCMO layer, and the other in which the magnetization is inhomogeneous such that the 
magnetization is suppressed close to the LCMO/YBCO interface.  For comparison of the 
two cases, we plot in Fig.~\ref{fig:Polarization} the measured neutron spin asymmetry 
$(R^{++}- R^{--})/(R^{++}+R^{--})$ together with the best fits for both the homogeneous 
and the inhomogeneous model.  Notice, that the sign of the spin asymmetry is opposite for 
the two model calculations at the position of the second superlattice Bragg peak 
(at 0.15~1/\AA).  The sign of the spin asymmetry is positive for the inhomogeneous case, 
while it is negative for the homogeneous case.  Experimentally the asymmetry integrated 
around the second superlattice peak ($q = 0.13$--0.17~1/\AA) is clearly positive: 
$A = 0.4\pm0.3$.  This shows unambiguously that the magnetization in the LCMO is 
reduced close to the interface with YBCO.  In addition the experimental spin asymmetry 
near the critical edge and first superlattice Bragg peak is significantly better fitted by the 
inhomogeneous model.  A direct comparison between the experimental polarized neutron 
reflectivities and the fits based on the inhomogeneous model is also shown in 
Fig.~\ref{fig:PNR}.  An even better fit for the neutron data could be obtained if the 
structural parameters were also varied.

Recently, Stahn {\em at al.}\ explained their neutron reflectometry data with two possible 
scenarios; one with a suppressed magnetization in the LCMO, similar to our case, and the 
other one where close to the interface a net magnetic moment is induced in the YBCO layer, 
which should be antiparallel to the LCMO moments.\cite{PSI}  We can exclude the latter 
possibility, since in our measurements, the reversal in sign of the spin asymmetry around the 
second Bragg peak is a smoking gun for the suppressed magnetization.  We can exclude an 
additionally induced net magnetization in the YBCO, either parallel or antiparallel to the 
magnetization in the LCMO.  Simulations for both cases always result in the wrong sign of 
the spin asymmetry at the second Bragg peak, similar to the homogenous model discussed 
previously.

\begin{figure}
\includegraphics[width=3.4in]{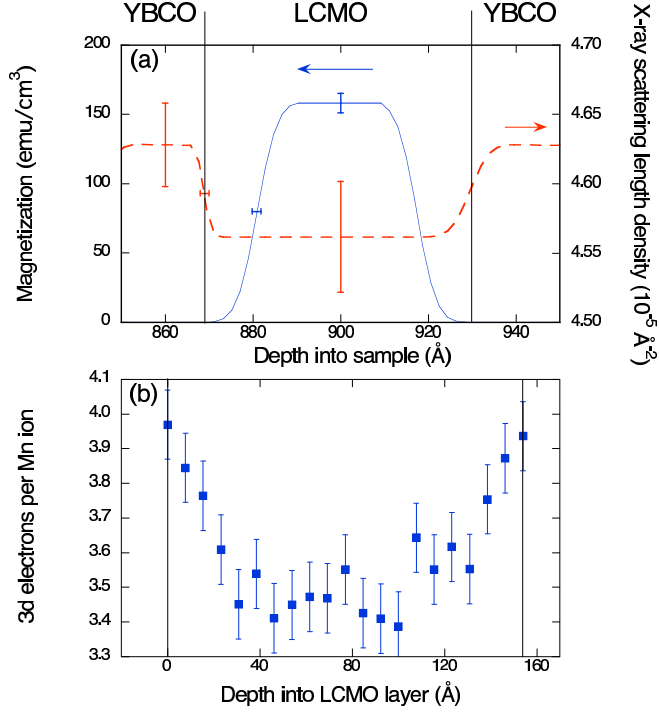}
\caption{(Color online) (a) Depth profile of the magnetization (from polarized neutron 
reflectometry) and chemical structure given by the X-ray scattering length density for the 
[LCMO 15 u.c./YBCO 3 u.c.] superlattice.  The vertical lines indicate the structural 
LCMO/YBCO interfaces.  Error bars for the model parameters of the fit are also indicated.  
(b)  Occupation of the Mn 3d band as determined from EELS spectroscopy as a function of 
distance across one LCMO layer for a [LCMO 40 u.c./YBCO 12 u.c.] superlattice.}
\label{fig:Profile}
\end{figure}

The structural profile obtained from the fit to the X-ray data, and the magnetization profile 
obtained from the fit to the neutron data are shown in detail in Fig.~\ref{fig:Profile}(a).  
The magnetic layer thickness is reduced compared to the structural layer thickness over a 
distance up to $12 \pm 1$~\AA\ away from the interface into the LCMO.  Furthermore, the 
magnetization in the middle of each LCMO layer ($158 \pm 7$~emu/cm$^3$) is only about 
1/3 of the bulk LCMO magnetization value.  This reduced value may be partly due to the 
elevated temperature of the measurement and partly due to a charge transfer between the 
YBCO and the LCMO layer as discussed below.

In order to further investigate the origin of the reduced magnetization we performed 
EELS spectroscopy with atomic resolution.  From the analysis of the Mn L edge at 640~eV 
the formal oxidation state of Mn in the LCMO can be 
determined\cite{Mn-core_edge,Mn-EELS}.  Figure~\ref{fig:Profile}(b) shows the 
variation of the Mn 3d band occupations across a single LCMO layer in a 
[LCMO 40 u.c./YBCO 12 u.c.] superlattice.  These measurements were averaged over a 
lateral distance parallel to the interface of 6~nm.  Based on the chosen chemical doping of 
the LCMO we anticipate 3.67 3d electrons per Mn ion, which is in good agreement with the 
occupation $3.62\pm0.18$ measured by EELS averaged over the whole LCMO layer.  
However, near the YBCO/LCMO interface a spatially inhomogeneous distribution was 
found.  The number of 3d electrons per formula unit was found to increase, and the increase 
scaled inversely with the magnetization suppression.  Interestingly, for close to four 3d 
electrons per Mn ion LCMO exhibits antiferromagnetic ordering.  Also, the EELS data show 
an overall decrease in the number of 3d electrons per Mn ion in the center of the layer.  A 
survey of other similarly prepared samples also found a general decrease of the electron 
occupation near the film center\cite{Maria-preprint}.  This suggests that there is a net 
charge transfer from the YBCO to the LCMO layer.  Recently infrared absorption also found 
a long-range charge transfer in YBCO/LCMO superlattices\cite{IR}.

It is useful to estimate in the LCMO layer the Thomas-Fermi screening 
length $\lambda_{TF} = 1/2\sqrt{a_0/n^{1/3}}$, which is the typical length scale 
for charge inhomogeneities.  Here $a_0$ is the Bohr radius and $n$ is the charge carrier 
density, which in manganites is typically $10^{19}$--$10^{22}$~cm$^{-3}$\cite{Ahn}.  
Using this range of charge carrier density, we obtain $\lambda_{TF} = 2$--6~\AA, which 
corresponds to a distance of one to two unit cells of the LCMO layers.  From the fit to the 
neutron data we obtain a $12 \pm 1$~\AA\ thick region over which the magnetization is 
suppressed within each LCMO layer.  The similarity of these length scales suggests that the 
suppression of the saturation magnetization may be due to a charge transfer across the 
interface.  Also notice that band bending at the interface might give rise to a longer distance 
for charge redistributions than estimated by $\lambda_{TF}$.   The continuously varying 
doping, as evidenced from the EELS measurements, may give rise to antiferromagnetic 
order close to the YBCO/LCMO interface, which again is consistent with the suppressed 
magnetization at the interface.  The LCMO recovers ferromagnetic order only once it is 
further away from the interface.  This interpretation of the data is also consistent with the 
recently observed exchange bias in YBCO/LCMO superlattices arising from exchange 
coupling between a ferromagnet and an 
antiferromagnet\cite{exchange_bias,exchange_bias2}.

The depressed magnetization at the interface may also provide an explanation for the long 
range proximity effect reported recently in YBCO/LCMO heterostructures.  Measurements of 
$T_c$ of F/S/F trilayers and superlattices as a function of the thickness of the F layer suggest 
that the order parameter penetrates distances up to 5~nm into the 
ferromagnet\cite{long-proximity}.  There should not be any ferromagnetic / (singlet) 
superconducting proximity effect if the LCMO were half metallic\cite{Andreev}.  
However, the suppressed magnetization at the interface and the depressed magnetization 
value within the LCMO layer suggest otherwise.  The exchange splitting $\Delta E_{ex}$ 
is connected to the magnetic moment ($\mu$) through an effective exchange integral 
$I_{eff}$ as $\Delta E_{ex} = \mu I_{eff}$.  The reduced moment may be reflecting a 
decreased exchange splitting thus providing a scenario for the penetration of the 
superconductivity into the ferromagnet\cite{transparency}.

In conclusion, we have shown unambiguously that the magnetization in LCMO layers is 
suppressed at the interface with YBCO.  The suppression of magnetization at the interface is 
correlated with an increased occupancy of electron charge at the Mn sites.  This suppression 
of magnetization may be a consequence of the redistribution of electric charges at the 
LCMO/YBCO interface. 

We would like to acknowledge stimulating discussions with H.-U.~Habermeier and 
J.~Chakalian.  This work was funded by MCYT MAT 2002-2642, Fundaci\'on Ram\'on 
Areces, CAM GR-MAT-0771/2004, and by the U.~S.\ DOE, BES under contracts 
W-31-109-ENG-38, W-7405-ENG-36, and DE-AC05-00OR22725.

\end{document}